\documentclass[reprint,showpacs,amsmath,amssymb,aps,prl,superscriptaddress]{revtex4-1}

\usepackage{graphicx}
\usepackage{dcolumn}
\usepackage{bm}
\usepackage{latexsym}
\usepackage{wasysym}
\usepackage{lineno}
\usepackage{stackrel}
\usepackage{color}

\makeatletter
\def\overlay#1#2{\mathpalette\@overlay{{#1}{#2}{\hfil}{\hfil}}}
\def\@overlay#1#2{\@@overlay#1#2}
\def\@@overlay#1#2#3#4#5{{%
   \def\overlaystyle{#1}%
   \setbox0=\hbox{\m@th$\overlaystyle#2$}%
   \setbox1=\hbox{\m@th$\overlaystyle#3$}%
   \ifdim \wd0<\wd1 \setbox2=\box1 \setbox1=\box0 \setbox0=\box2\fi
   \rlap{\hbox to\wd0{#4\box1\relax#5}}\box0%
}}
\makeatother

\newcommand{\SuperCirc}{\overlay{}{\circlearrowleft}}

\makeatletter
\newlength\Cover@height
\newlength\Cover@depth
\newlength\Cover@width
\newcommand{\Cover}[2][]{%
  \settoheight{\Cover@height}{$#2$}
  \settodepth{\Cover@depth}{$#2$}
  \settowidth{\Cover@width}{$#2$}
  \def\Cover@arg{#1}
  \raisebox{.5ex}{\raisebox{\Cover@height}{\rlap{%
  $\SuperCirc$}}}
  #2
}
\makeatother

\newcommand{\CirPol}{\Cover{\gamma}}

\begin{document}

%
%

\title{Production of Highly-Polarized Positrons Using Polarized Electrons at MeV Energies}

\affiliation{Thomas Jefferson National Accelerator Facility, Newport News, VA 23606, USA}
\affiliation{LPSC, Universit\'e Grenoble-Alpes, CNRS/IN2P3, 38026 Grenoble, France}
\affiliation{Hampton University, Hampton, VA 23668, USA}
\affiliation{Idaho State University, Pocatello, ID 83209, USA}
\affiliation{LAL, Universit\'e Paris-Sud \& Universit\'e Paris-Saclay, CNRS/IN2P3, 91898 Orsay, France}
\affiliation{IPN, Universit\'e Paris-Sud \& Universit\'e Paris-Saclay, CNRS/IN2P3, 91406 Orsay, France}
\affiliation{North Carolina Central University, Durham, NC 27707, USA}
\affiliation{Old Dominion University, Norfolk, VA 23529, USA}
\affiliation{Rutgers, The State University of New Jersey, Piscataway, NJ 08854, USA}
\affiliation{The College of William \& Mary, Williamsburg, VA 23187, USA}
\affiliation{University of Virginia, Charlottesville, VA 22901, USA}
\affiliation{Virginia Polytechnique Institut and State University, Blacksburg, VA 24061, USA}
\affiliation{University of New Mexico, Albuquerque, NM 87131, USA}
\affiliation{The George Washington University, Washington, DC 20052, USA}

\author{D.~Abbott}         
\affiliation{Thomas Jefferson National Accelerator Facility, Newport News, VA 23606, USA}

\author{P.~Adderley}       
\affiliation{Thomas Jefferson National Accelerator Facility, Newport News, VA 23606, USA}

\author{A.~Adeyemi}
\affiliation{Hampton University, Hampton, VA 23668, USA}

\author{P.~Aguilera}
\affiliation{Thomas Jefferson National Accelerator Facility, Newport News, VA 23606, USA}

\author{M.~Ali}  
\affiliation{Thomas Jefferson National Accelerator Facility, Newport News, VA 23606, USA}

\author{H.~Areti}
\affiliation{Thomas Jefferson National Accelerator Facility, Newport News, VA 23606, USA}

\author{M.~Baylac}
\affiliation{LPSC, Universit\'e Grenoble-Alpes, CNRS/IN2P3, 38026 Grenoble, France}

\author{J.~Benesch}
\affiliation{Thomas Jefferson National Accelerator Facility, Newport News, VA 23606, USA}

\author{G.~Bosson}
\affiliation{LPSC, Universit\'e Grenoble-Alpes, CNRS/IN2P3, 38026 Grenoble, France}

\author{B.~Cade}
\affiliation{Thomas Jefferson National Accelerator Facility, Newport News, VA 23606, USA}

\author{A.~Camsonne}
\affiliation{Thomas Jefferson National Accelerator Facility, Newport News, VA 23606, USA}

\author{L.S.~Cardman}
\affiliation{Thomas Jefferson National Accelerator Facility, Newport News, VA 23606, USA}

\author{J.~Clark}
\affiliation{Thomas Jefferson National Accelerator Facility, Newport News, VA 23606, USA}

\author{P.~Cole}
\affiliation{Idaho State University, Pocatello, ID 83209, USA}

\author{S.~Covert}
\affiliation{Thomas Jefferson National Accelerator Facility, Newport News, VA 23606, USA}

\author{C.~Cuevas}
\affiliation{Thomas Jefferson National Accelerator Facility, Newport News, VA 23606, USA}

\author{O.~Dadoun}
\affiliation{LAL, Universit\'e Paris-Sud \& Universit\'e Paris-Saclay, CNRS/IN2P3, 91898 Orsay, France}

\author{D.~Dale}
\affiliation{Idaho State University, Pocatello, ID 83209, USA}

\author{H.~Dong}
\affiliation{Thomas Jefferson National Accelerator Facility, Newport News, VA 23606, USA}

\author{J.~Dumas}
\affiliation{Thomas Jefferson National Accelerator Facility, Newport News, VA 23606, USA}
\affiliation{LPSC, Universit\'e Grenoble-Alpes, CNRS/IN2P3, 38026 Grenoble, France}

\author{E.~Fanchini}
\affiliation{LPSC, Universit\'e Grenoble-Alpes, CNRS/IN2P3, 38026 Grenoble, France}

\author{T.~Forest}
\affiliation{Idaho State University, Pocatello, ID 83209, USA}

\author{E.~Forman}
\affiliation{Thomas Jefferson National Accelerator Facility, Newport News, VA 23606, USA}

\author{A.~Freyberger}
\affiliation{Thomas Jefferson National Accelerator Facility, Newport News, VA 23606, USA}

\author{E.~Froidefond}
\affiliation{LPSC, Universit\'e Grenoble-Alpes, CNRS/IN2P3, 38026 Grenoble, France}

\author{S.~Golge}
\affiliation{North Carolina Central University, Durham, NC 27707, USA}

\author{J.~Grames}
\affiliation{Thomas Jefferson National Accelerator Facility, Newport News, VA 23606, USA}

\author{P.~Gu\`eye}
\affiliation{Hampton University, Hampton, VA 23668, USA}

\author{J.~Hansknecht}
\affiliation{Thomas Jefferson National Accelerator Facility, Newport News, VA 23606, USA}

\author{P.~Harrell}
\affiliation{Thomas Jefferson National Accelerator Facility, Newport News, VA 23606, USA}

\author{J.~Hoskins}
\affiliation{The College of William \& Mary, Williamsburg, VA 23187, USA}

\author{C.~Hyde}
\affiliation{Old Dominion University, Norfolk, VA 23529, USA}

\author{B.~Josey}
\affiliation{University of New Mexico, Albuquerque, NM 87131, USA}

\author{R.~Kazimi}
\affiliation{Thomas Jefferson National Accelerator Facility, Newport News, VA 23606, USA}

\author{Y.~Kim}
\affiliation{Thomas Jefferson National Accelerator Facility, Newport News, VA 23606, USA}
\affiliation{Idaho State University, Pocatello, ID 83209, USA}

\author{D.~Machie}
\affiliation{Thomas Jefferson National Accelerator Facility, Newport News, VA 23606, USA}

\author{K.~Mahoney}
\affiliation{Thomas Jefferson National Accelerator Facility, Newport News, VA 23606, USA}

\author{R.~Mammei}
\affiliation{Thomas Jefferson National Accelerator Facility, Newport News, VA 23606, USA}

\author{M.~Marton}
\affiliation{LPSC, Universit\'e Grenoble-Alpes, CNRS/IN2P3, 38026 Grenoble, France}

\author{J.~McCarter}
\affiliation{University of Virginia, Charlottesville, VA 22901, USA}

\author{M.~McCaughan}
\affiliation{Thomas Jefferson National Accelerator Facility, Newport News, VA 23606, USA}

\author{M.~McHugh} 
\affiliation{The George Washington University, Washington, DC 20052, USA}

\author{D.~McNulty} 
\affiliation{Idaho State University, Pocatello, ID 83209, USA}

\author{K.E.~Mesick}     
\affiliation{Rutgers, The State University of New Jersey, Piscataway, NJ 08854, USA}

\author{T.~Michaelides}  
\affiliation{Thomas Jefferson National Accelerator Facility, Newport News, VA 23606, USA}

\author{R.~Michaels} 
\affiliation{Thomas Jefferson National Accelerator Facility, Newport News, VA 23606, USA}

\author{B.~Moffit}  
\affiliation{Thomas Jefferson National Accelerator Facility, Newport News, VA 23606, USA}

\author{D.~Moser}  
\affiliation{Thomas Jefferson National Accelerator Facility, Newport News, VA 23606, USA}

\author{C.~Mu\~noz~Camacho}
\affiliation{IPN, Universit\'e Paris-Sud \& Universit\'e Paris-Saclay, CNRS/IN2P3, 91406 Orsay, France}

\author{J.-F.~Muraz}  
\affiliation{LPSC, Universit\'e Grenoble-Alpes, CNRS/IN2P3, 38026 Grenoble, France}

\author{A.~Opper}  
\affiliation{The George Washington University, Washington, DC 20052, USA}

\author{M.~Poelker}    
\affiliation{Thomas Jefferson National Accelerator Facility, Newport News, VA 23606, USA}

\author{J.-S.~R\'eal}
\affiliation{LPSC, Universit\'e Grenoble-Alpes, CNRS/IN2P3, 38026 Grenoble, France}

\author{L.~Richardson}    
\affiliation{Thomas Jefferson National Accelerator Facility, Newport News, VA 23606, USA}

\author{S.~Setiniyaz}     
\affiliation{Idaho State University, Pocatello, ID 83209, USA}

\author{M.~Stutzman}     
\affiliation{Thomas Jefferson National Accelerator Facility, Newport News, VA 23606, USA}

\author{R.~Suleiman}      
\affiliation{Thomas Jefferson National Accelerator Facility, Newport News, VA 23606, USA}

\author{C.~Tennant}    
\affiliation{Thomas Jefferson National Accelerator Facility, Newport News, VA 23606, USA}

\author{C.~Tsai}       
\affiliation{Virginia Polytechnique Institut and State University, Blacksburg, VA 24061, USA}

\author{D.~Turner}      
\affiliation{Thomas Jefferson National Accelerator Facility, Newport News, VA 23606, USA}

\author{M.~Ungaro}      
\affiliation{Thomas Jefferson National Accelerator Facility, Newport News, VA 23606, USA}

\author{A.~Variola} 
\affiliation{LAL, Universit\'e Paris-Sud \& Universit\'e Paris-Saclay, CNRS/IN2P3, 91898 Orsay, France}

\author{E.~Voutier}   
\affiliation{LPSC, Universit\'e Grenoble-Alpes, CNRS/IN2P3, 38026 Grenoble, France}
\affiliation{IPN, Universit\'e Paris-Sud \& Universit\'e Paris-Saclay, CNRS/IN2P3, 91406 Orsay, France}

\author{Y.~Wang}      
\affiliation{Thomas Jefferson National Accelerator Facility, Newport News, VA 23606, USA}

\author{Y.~Zhang}	      
\affiliation{Rutgers, The State University of New Jersey, Piscataway, NJ 08854, USA}

\collaboration{PEPPo Collaboration}

\makeatletter
\global\@specialpagefalse
\def\@oddhead{
\hfill {D.~Abbott {\it et al.}, Production of highly-polarized positrons using polarized electrons at MeV energies}
}
\let\@evenhead\@oddhead
\def\@oddfoot{\reset@font\rm\hfill \thepage\hfill
} \let\@evenfoot\@oddfoot
\makeatother

\begin{abstract}

The Polarized Electrons for Polarized Positrons experiment at the injector of the Continuous Electron Beam Accelerator 
Facility has demonstrated for the first time the efficient transfer of polarization from electrons to positrons produced 
by the polarized bremsstrahlung radiation induced by a polarized electron beam in a high-$Z$ target. Positron polarization 
up to 82\% have been measured for an initial electron beam momentum of 8.19~MeV/$c$, limited only by the electron 
beam polarization. This technique extends polarized positron capabilities from GeV to MeV electron beams, and opens access 
to polarized positron beam physics to a wide community.

\end{abstract}


\maketitle

%
%
%
%

Positron beams, both polarized and unpolarized, with energies ranging from a few eV to hundreds of GeV are unique tools for 
the study of the physical world. For energies up to several hundred keV, they allow the study of surface magnetization 
properties of materials~\cite{Gid82} and their inner structural defects~\cite{Kra99}. In the several to tens of GeV energy range, 
they provide the complementary experimental observables essential for an unambiguous determination of the structure of the 
nucleon~\cite{Vou14}. In the several hundreds of GeV energy range, they are considered essential for the next generation of 
experiments that will search for physics beyond the Standard Model~\cite{Beh13}. Unfortunately, the creation of polarized positron 
beams is especially difficult. Radioactive sources can be used for low energy positrons~\cite{Zit79}, but the flux is restricted. 
Storage or damping rings can be used at high energy, taking advantage of the self-polarizing Sokolov-Ternov effect~\cite{Sok64}, 
however, this approach is generally not suitable for external beams and continuous wave facilities.

Recent schemes for polarized positron production at such proposed facilities rely on the polarization transfer in the $e^+e^-$-pair 
creation process from circularly polarized photons~\cite{{Ols59},{Kur10}}, but use different methods to produce the polarized 
photons. Two techniques have been investigated successfuly: the Compton backscattering of polarized laser light from a GeV 
unpolarized electron beam~\cite{Omo06}, and the synchrotron radiation of a multi-GeV unpolarized electron beam travelling within 
a helical undulator~\cite{Ale08}. Both demonstration experiments reported high positron polarization, confirming the efficiency of 
the pair production process for producing a polarized positron beam. However, these techniques require high energy electron beams 
and challenging technologies that limit their range of application.

A new approach, which we refer to as the Polarized Electrons for Polarized Positrons (PEPPo) concept~\cite{{Bes96},{Pot97}}, 
has been investigated at the Continuous Electron Beam Accelerator Facility (CEBAF) of the Thomas Jefferson National Accelerator 
Facility (JLab). Taking advantage of advances in high polarization, high intensity electron sources~\cite{Add10}, it exploits the polarized 
photons generated by the bremsstrahlung radiation of low energy longitudinally polarized electrons within a high-$Z$ target to produce 
polarized $e^+e^-$-pairs. It is expected that the PEPPo concept can be developed efficiently with a low energy ($\sim$5-100~MeV/$c$), 
high intensity ($\sim$mA), and high polarization ($>$80\%) electron beam driver, opening access to polarized positron beams to a 
wide community.

%
%
%
 
\begin{figure}[t]
\includegraphics[width=0.985\linewidth,angle=0]{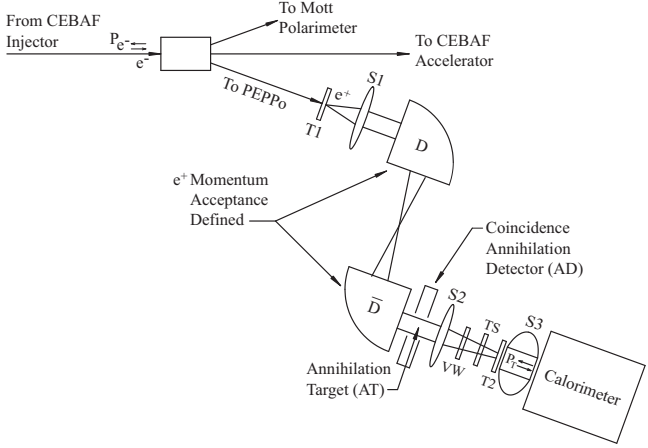}
\vspace*{-10pt}
\caption{\label{PEPPo-line} Schematic of the PEPPo line and apparatus illustrating the principle of operation of the experiment 
based on the processes sequence $\protect\overrightarrow{e^-} \stackrel{\mbox{\tiny{T1}}}{\rightarrow} \CirPol 
\stackrel{\mbox{\tiny{T1}}}{\rightarrow} \protect\overrightarrow{e^+} \stackrel{\mbox{\tiny{T2}}}{\rightarrow} \CirPol 
\stackrel{\mbox{\tiny{S3}}}{\rightarrow} \gamma$ described in the text. The setup footprint is about 3$\times$1.5~m$^2$.}
\end{figure}
The present experiment~\cite{Gra11} was designed to evaluate the PEPPo concept by measuring the 
polarization transfer from a primary electron beam to the produced positrons. A new beam line (Fig.~\ref{PEPPo-line}) was constructed 
at the CEBAF injector~\cite{Kaz04} where polarized electrons up to 8.19~MeV/$c$ were transported to a 1~mm thick tungsten 
positron production target (T1) followed by a positron collection, selection, and characterization system~\cite{Ale09}. Longitudinally 
polarized electrons interacting in T1 radiate elliptically polarized photons whose circular component ($P_{\gamma}$) is 
proportional to the electron beam polarization ($P_e$). Within the same target, the polarized photons produce polarized 
$e^+e^-$-pairs with perpendicular ($P_{\perp}$) and longitudinal ($P_{\parallel}$) polarization components both proportional 
to $P_{\gamma}$ and therefore $P_e$. The azimuthal symmetry causes $P_{\perp}$ to vanish resulting in longitudinally polarized secondary 
positrons. Immediately after T1, a short focal length solenoid (S1) collects the positrons into a combined function spectrometer 
($\mathrm{D}\overline{\mathrm{D}}$) composed of two 90$^{\circ}$ dipoles that select positron momentum. The exiting positrons can either 
be detected at a positron diagnostic (AT+AD) or refocused by a second solenoid (S2) through a vacuum window (VW) to a Compton 
transmission polarimeter. Retracting T1, the known electron beam was additionally transported to T2 to calibrate the polarimeter 
analyzing power. 

This polarimeter~\cite{Ale09} begins with a 2~mm densimet (90.5\%W/7\%Ni/2.5\%Cu) conversion target (T2) followed by a 7.5~cm 
long, 5~cm diameter iron cylinder centered in a solenoid (S3) that saturates and polarizes it. The average longitudinal polarization 
was measured to be $\overline{P_T} = 7.06\pm0.09$\%, in very good agreement with the previously reported value~\cite{Ale09}. An 
electromagnetic calorimeter with 9 CsI crystals (6$\times$6$\times$28~cm$^3$) arranged in a 3$\times$3-array is placed at the exit of 
the polarimeter solenoid. Polarized positrons convert at T2 via bremsstrahlung and annihilation processes into polarized photons with polarization orientation and magnitude that depend on the positron polarization. Because of the polarization dependence of the Compton process, the number of photons passing through the iron core and subsequently detected by the CsI-array depends on the relative 
orientation of the photon and iron core polarizations. By reversing the sign of the positron polarization (via the electron beam 
helicity) or the target polarization (via S3 polarity), one measures the experimental Compton asymmetry
\begin{equation}
A_C^p = P_{\parallel} \, \overline{P_T} \, A_p = \epsilon_P \, P_e \, \overline{P_T} \, A_p \label{eq:ComAs}
\end{equation}
where $A_p$ is the positron analyzing power of the polarimeter and $\epsilon_P$ is the electron-to-positron polarization transfer efficiency. Knowing $\overline{P_T}$, $P_e$, $A_p$ and measuring $A_C^p$ provide a measurement of $P_{\parallel}$ and $\epsilon_P$.

%
%
%

\begin{figure}[b!]
\includegraphics[width=0.80\linewidth]{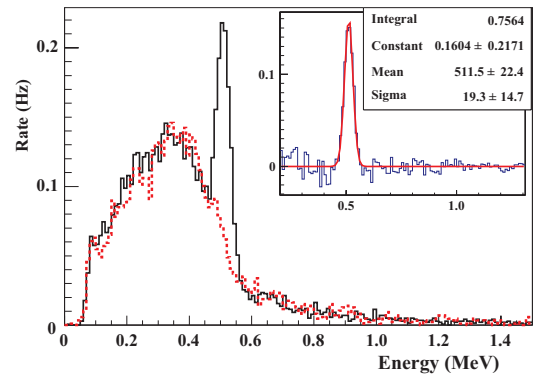}
\vspace*{-10pt}
\caption{\label{AnniSig} (Color online) Measured energy spectra in one of the annihilation detectors (AD) once T1 is inserted in 
(full line) or removed from (dash line) the electron beam path; the top right corner shows the difference between these spectra.}
\end{figure}
PEPPo used a polarized electron beam of $p_e$=8.19$\pm$0.04~MeV/$c$ to measure the momentum dependence of $\epsilon_P$ over the 
positron momentum range of 3.07 to 6.25~MeV/$c$. The magnetic beam line and polarimeter were first calibrated using electron beams 
of precisely measured polarization and with momenta adjusted to match the positron momenta to be studied. Only the polarity of the 
spectrometer was reversed when measuring positrons instead of electrons. The uncertainties on the positron momenta (Tab.~\ref{TabRes}) 
arise from the electron calibration procedure and hysteresis effects under polarity reversal for positron collection. The 
experimental values of S1, $\mathrm{D}\overline{\mathrm{D}}$, and S2 currents agree well with those determined by a 
GEANT4~\cite{Ago03} model of the experiment using magnetic fields modeled with OPERA-3D~\cite{Ben15}. 
 
The polarization of the electron beam, $P_e$, was measured to be $85.2\pm0.6\pm0.7$\% with a Mott polarimeter~\cite{Gra13}. The 
first uncertainty is statistical and the second is the total systematic uncertainty associated with the theoretical and experimental 
determination of the Mott analyzing power. 

The AD diagnostic is used to demonstrate the presence of positrons exiting the $\mathrm{D}\overline{\mathrm{D}}$ spectrometer. 
When interacting with an insertable chromium oxide target (AT in Fig.~\ref{PEPPo-line}) at the spectrometer exit, positrons annihilate 
into two back-to-back 511~keV photons (Fig.~\ref{AnniSig}) that are detected by a pair of NaI detectors (AD in Fig.~\ref{PEPPo-line}). 

%
%
%
 
 The polarimeter's CsI crystal array was read out by photomultipliers (PMT). The effective gain of each crystal was calibrated prior 
 to beam exposure with $^{137}$Cs and $^{22}$Na radioactive sources, and monitored during data taking by controlling the position of 
 the 511~keV peak produced by the annihilation of positrons created in the iron core. This method insures a robust and stable energy 
 measurement, intrinsically corrected for possible radiation damage or PMT-aging effects. A positron trigger was formed from a 
 coincidence between the central crystal and a 1~mm thick scintillator (TS in Fig.~\ref{PEPPo-line}) placed between the beam 
 line vacuum exit window and T2; it constitutes an effective charged particle trigger that considerably reduces the neutral background  
 in the crystal array. 
 
 The electronic readout operated in two modes: single event mode; and integrated mode in which the PMT signal from the crystal was 
 integrated over the total time associated with a fixed beam polarization orientation (helicity gate). This mode was used in high 
 rate background-free situations.  
  
\begin{figure}[t]
\includegraphics[width=0.985\linewidth]{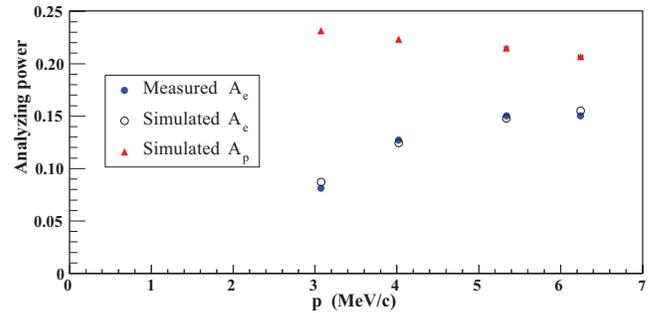}
\vspace*{-10pt}
\caption{\label{AnaPow} (Color online) Electron and positron analyzing powers of the central crystal of the polarimeter. 
Statistical uncertainties were combined quadratically with systematic uncertainties taken from $P_e$, $\overline{P_T}$, and $A_C^e$ 
to determine acutal error bars.}
\end{figure}

 Electron calibration data were recorded in integrated mode. The comparison of the total energy deposited ($E^{\pm}$) as the electron 
 beam helicity is toggled ($\pm$) at 30~Hz is formed and defines the experimental asymmetry. Occasional reversal of the sign of the  
 experimental asymmetry was applied to suppress systematic effects of target polarization (reversing S3   polarity) or electron 
 polarization (reversing polarization of the source laser with a half-wave plate). The results were combined statistically to provide 
 the actual Compton asymmetries $A_C^e$ for electrons 
 \begin{equation}
 A_C^e = P_e \, \overline{P_T} \, A_e  \label{eq:StaC} 
 \end{equation}
 where $A_e$ is the electron analyzing power of the polarimeter. Experimental values reported in Tab.~\ref{TabRes} feature high 
 statistical accuracy ($<1$\%) and comparable systematic errors originating from the determination of the pedestal signal. Since 
 the beam and target polarizations are known, these constitute measurements of $A_e$ (Eq.~\ref{eq:StaC}). The experimental analyzing 
 power increases with electron momentum, as expected (Fig.~\ref{AnaPow}).

%
%
%
 
 Positron data were recorded on an event-by-event basis and, because of the trigger configuration, involve only the central crystal. 
 The experimental information consists of the energy deposited in that crystal and the coincidence time ($t_c$) between TS and the 
 crystal. The energy yield was determined for each helicity state by summing the energy deposited for each event occuring during the  
 corresponding helicity gate, normalized by the beam charge associated with that helicity state and corrected for electronic and data  
 acquisition dead-time measured with specific helicity-gated scalers. Data were further corrected for random coincidences by an analysis 
 of the time spectra. The statistical combination of the data for each S3 polarity and helicity configuration provides the Compton 
 asymmetry $A_C^p$ (Eq.~\ref{eq:ComAs}). Tab.~\ref{TabRes} reports experimental asymmetries and  uncertainties for each positron momentum, 
 integrating over energy deposited above 511~keV. The main sources of systematics originate from the energy calibration procedure, the 
 random subtraction method, and the selection of coincidence events. They are quadratically combined to yield Tab.~\ref{TabRes} values, 
 whose dominant contributions are from the subtraction of random coincidence events.   

%
%
%

\begingroup
\squeezetable
\begin{table*}[t!]
\caption{\label{TabRes} PEPPo asymmetry measurements and polarization data at the central crystal.}
\begin{ruledtabular}
\begin{tabular}{cc||ccc|ccc||ccc|ccc}
\multicolumn{2}{c||}{Momentum} & \multicolumn{6}{c||}{Experimental asymmetries} & \multicolumn{6}{c}{Polarization data} \\
 $p$ & $\delta p$ & $A_C^e$ & $\delta A_C^{e \, Sta.}$ & $\delta A_C^{e \, Sys.}$ & $A_C^p$ & $\delta A_C^{p \, Sta.}$ & 
 $\delta A_C^{p \, Sys.}$ & $\epsilon_P$ & $\delta \epsilon_P^{Sta.}$ & $\delta \epsilon_P^{Sys.}$ & $P_{\parallel}$ & 
 $\delta P_{\parallel}^{Sta.}$ & $\delta P_{\parallel}^{Sys.}$ \\
 (MeV/$c$) & (MeV/$c$) & ($\permil$) & ($\permil$) & ($\permil$) & ($\permil$) & ($\permil$) & ($\permil$) & (\%) & (\%) 
 & (\%) & (\%) & (\%) & (\%) \\ \hline
 3.07 & 0.02 & 4.89 & 0.03 & 0.07 & 7.03 & 1.06           & 0.17           & 50.4 & 7.6 & 1.4 & 43.0 & 6.5 & 1.1 \\
 4.02 & 0.03 & 7.65 & 0.05 & 0.07 & 8.71 & 0.49           & 0.13           & 64.8 & 3.7 & 1.4 & 55.2 & 3.2 & 1.1 \\
 5.34 & 0.03 & 9.03 & 0.03 & 0.03 & 11.3 & 0.5\phantom{0} & 0.1\phantom{0} & 87.4 & 3.8 & 1.6 & 74.4 & 3.2 & 1.2 \\
 6.25 & 0.04 & 9.04 & 0.04 & 0.04 & 12.0 & 0.4\phantom{0} & 0.1\phantom{0} & 96.5 & 3.2 & 1.7 & 82.2 & 2.7 & 1.3 \\
\end{tabular}
\end{ruledtabular}
\end{table*}
\endgroup
\begin{figure}[h!]
\includegraphics[width=0.985\linewidth]{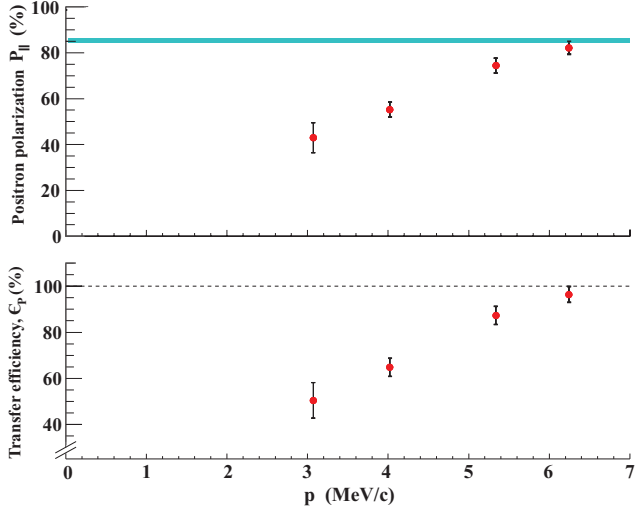}
\vspace*{-10pt}
\caption{\label{PolPos} (Color online) PEPPo measurements of the positron polarization (top panel) and polarization transfer 
efficiency (bottom panel); statistics and systematics are reported for each point, and the shaded area indicates the electron 
beam polarization.}
\end{figure}

The PEPPo beam line, magnet fields, and detection system was modeled using GEANT4, taking advantage of a previous 
implementation of polarized electromagnetic processes~\cite{{Dol06},{Dum09}}. The calibration of the analyzing power of the 
polarimeter relies on the comparison between experimental and simulated electron asymmetries. It allowed us to benchmark the 
GEANT4 physics packages and resolve related systematic uncertainties within the limits of the measurement accuracy. The excellent 
agreement between electron measurements and simulations (Fig.~\ref{AnaPow}) indicates an accurate understanding of the beam line 
optics and the quality of the operation of the polarimeter. Finally, the analyzing power of the polarimeter for positrons may be 
directly simulated (Fig.~\ref{AnaPow}). Conditions for the simulations are guided by the optics of the beam line leading to the 
vacuum window and bounded by the actual largest $e^+$ beam size that may reach T2. The combination of the supplementary $e^+$-to-$\gamma$ annihilation conversion process together with the minimum energy deposited requirement (511~keV) leads to larger analyzing power 
for positrons than electrons. The energy cut effect is strong at low $e^+$ momenta where it removes a significant part of the energy 
spectra acting as a dilution of the polarization sensitivity.
 
%
%
%

The positron longitudinal polarization $P_{\parallel}$ and the polarization transfer efficiency $\epsilon_P$ as obtained from 
Eq.~\ref{eq:ComAs} are reported in Tab.~\ref{TabRes} and Fig.~\ref{PolPos}. These data show large positron polarization 
($P_{\parallel} > 40$\%) and polarization transfer efficiency ($\epsilon_P > 50$\%) over the explored momentum range. As a consequence
of ionization energy loss in the production target, large polarization transfer ($> 80$\%) is observed down to positron momenta as low 
as 60\% of the initial beam momentum. The bremsstrahlung of longitudinally polarized electrons is therefore demonstrated as an efficient 
process to generate longitudinally polarized positrons. The $e^+$ production efficiency deduced from the analysis of the photon rates at 
AD is $\sim$10$^{-6}$, in agreement with expectations~\cite{Gra11} and optical properties. 

%
%
%
This experiment successfully demonstrated the PEPPo concept by measuring longitudinal positron polarization up to 82\% corresponding 
to a polarization transfer approaching 100\% from 8.19~MeV/$c$ polarized electrons. These results expand the possibilities for the 
production of high intensity polarized positron beams from GeV to MeV accelerators.

Exploiting these conditions opens a large field of applications ranging from thermal polarized positron facilities to high energy  
colliders. These results can be extrapolated to any initial electron beam energy above the pair production threshold, depending on 
the desired positron flux and polarization. For each polarized positron source designed using the PEPPo concept, it will be essential 
to optimize the figure-of-merit incorporating the longitudinal and transverse emittance requirements of the application. For an 
accelerator like CEBAF, using the current polarized electron source, preliminary studies indicate that such an optimization would 
result in a polarized positron energy about half of the electron beam energy, a polarization transfer efficiency about 
75\%, and positron beam efficiencies about $10^{-4} \, e^+/e^-$ for initial beam momentum $\sim$100~MeV/$c$~\cite{Dum11}. 

%
%
%

\begin{acknowledgments}

We are deeply grateful to the SLAC E-166 Collaboration, particularly K.~Laihem, K.~McDonald, S.~Riemann, A.~Sch\"alicke, P.~Sch\"uler, 
J.~Sheppard and A.~Stahl for loan of fundamental equipment parts and support in GEANT4 modeling. We also thank N.~Smirnov for delivery 
of critical hardware. This work was supported in part by the U.S. Department of Energy, the French Centre National de la Recherche 
Scientifique, and the International Linear Collider R\&D program. Jefferson Science Associates operates the Thomas Jefferson National 
Accelerator Facility under DOE contract DE-AC05-06OR23177. 

\end{acknowledgments}

%
%

\bibliography{PEPPo-prl}

%
%

\end{document}